\documentclass[letterpaper,12pt]{JHEP3}
\usepackage{graphicx}
\usepackage{amsmath}
\usepackage{amsfonts}
\usepackage{amssymb}
\def\beq{\begin{equation}}
\def\eeq{\end{equation}}
\def\bea{\begin{eqnarray}}
\def\eea{\end{eqnarray}}
\def\lsim{\mathrel{\raise.3ex\hbox{$<$\kern-.75em\lower1ex\hbox{$\sim$}}}}
\def\gsim{\mathrel{\raise.3ex\hbox{$>$\kern-.75em\lower1ex\hbox{$\sim$}}}}
\def\ifmath#1{\relax\ifmmode #1\else $#1$\fi}

\def\gev{~{\mbox{GeV}}}

    \def\fillboxx#1#2{\hbox to #1{\vbox to #2{\vfil}\hfil}   }

\def\gev{~{\rm GeV}}

\def\cnone{\wt\chi^0_1}

\def\cntwo{\wt\chi^0_2}

\def\wt{\widetilde}

\newcommand{ \slashchar }[1]{\setbox0=\hbox{$#1$}   
   \dimen0=\wd0                                     
   \setbox1=\hbox{/} \dimen1=\wd1                   
   \ifdim\dimen0>\dimen1                            
      \rlap{\hbox to \dimen0{\hfil/\hfil}}          
      #1                                            
   \else                                            
      \rlap{\hbox to \dimen1{\hfil$#1$\hfil}}       
      /                                             
   \fi}     
\def\misspt{\slashchar p_T}
\title{ Minimal Kinematic Constraints and $m_{T2}$}

\author{Hsin-Chia Cheng and Zhenyu Han \\
Department of Physics, University of California, Davis, CA 95616\\
 E-mail: \email{cheng@physics.ucdavis.edu, zhenyuhan@physics.ucdavis.edu}}

\abstract{ We clarify the relation between the variable $m_{T2}$ and
  the method of kinematic constraints, both of which can be used
  for mass determination in events with two missing (dark matter) particles at
  hadron colliders. We identify a set of minimal kinematic
  constraints, including the mass shell conditions for the missing
  particles and their mother particles, as well as the constraint from
  the measured missing transverse momentum. We show that $m_{T2}$ is
  the boundary of the mass region consistent with the minimal
  constraints. From this point of view, we also obtained a more
  efficient algorithm for calculating $m_{T2}$. When more constraints 
  are available in the events, we can develop more sophisticated mass 
  determination methods starting from the $m_{T2}$ constraint. 
  In particular, we discuss cases when each decay chain
  contains two visible particles.
}

\begin{document}

\section{Introduction}
\label{sec:introduction}
Many extensions beyond the standard model (SM) predict a dark matter
candidate, that is, a massive stable particle interacting weakly with
the SM particles. These include the minimal supersymmetric standard
model (MSSM) with R-parity, universal extra dimension (UED) models
\cite{ued} with KK-parity, little Higgs models  with T-parity (LHT)
\cite{lht} and so forth. A common feature of these models is that they
all possess an exact parity, under which the SM particles are even and
some new physical states are odd. Therefore the lightest parity-odd
particle is stable and plays the role of the dark matter candidate. At
a collider, the parity odd particles must be produced in pairs. Each
of them will then go through cascade decays ending at the stable
particle. Because it is weakly interacting, the stable particle will
escape the detector without being detected, leaving missing energy
signals.

The Large Hadron Collider (LHC) will start collisions and collecting data
very soon. Once large missing energy signals are detected at the LHC,
it is crucial to investigate the properties of the new particles involved in
the events, in order to identify the underlying theory, as well as to
determine if the invisible particle is a viable dark matter
candidate. Of particular interest are the masses of the new particles,
including the invisible one. Due to the fact that there are always two
or more invisible particles in each of such events, mass determination
will be a difficult task. For a hadron collider, the total momentum in
the beam direction is not measured, making it even more challenging. 

Distributions of some simple kinematic variables such as $\slashchar
p_T$, $E_T$ and $M_{eff}$ \cite{meff} have been used to give estimates
of the masses of the new particles.  However, these variables are
mostly sensitive to the mass differences of the new particles, instead
of the absolute mass scale. On the other hand, the total production
cross section and the full likelihood method require knowledge of the
underlying physics such as the matrix elements, and hence are
model-dependent. Measurements of the new particle properties should be
the first step towards uncovering the underlying theory instead of the
other way around. It is therefore desirable to be able to determine
masses in a model-independent way by using only
kinematics. Traditionally, this has been done by looking for the
edges/endpoints of various invariant mass distributions of the visible
particles~\cite{edge}. The positions of these edges/endpoints are
functions of the masses of the particles involved in the decay
chains. If the decay chains are long enough, there may be enough
independent invariant mass endpoints involving the visible SM
particles, which allows one to reverse the relations to obtain the
masses. Nevertheless, this method applies to individual decay chains
and can only work for the long decay chains  (4 or more on-shell
particles in a decay chain). It does not utilize all information in
the events such as the measured missing transverse
momentum. Consequently, a large number of events are required to 
distinguish and measure all edges/endpoints in order to achieve a
reasonable determination of the masses. It is therefore important to
develop new mass determination techniques which are more powerful and
can also be applied to shorter decay chains for events involving
invisible particles.  

Recently, along this direction, two kinds of mass determination
techniques have been proposed. One of them utilizes the ``kinematic
constraints'' \cite{Kawagoe:2004rz, Cheng:2007xv,
  Cheng:2008mg}. Assuming that  the event topology is known, one can
try to reconstruct the kinematics event by event, by imposing the
mass-shell constraints and the constraint from the measured missing
transverse momentum, $\misspt$. Depending on how many constraints that
we can impose, the detailed methods can be different. In
Ref.~\cite{Cheng:2007xv}, the authors considered events with two
identical decay chains, each containing two visible
particles. Assuming that all intermediate particles are on-shell, the
two invisible particles' momenta can be solved for  given trial
masses. Requiring the solutions to be physical, one can determine the
masses by examining the number of solvable events for all possible
trial masses. Another method requires longer decay chains and
therefore more constraints. It is then possible to combine multiple
events and solve directly for the masses and the momenta, without
assuming any trial masses~\cite{Cheng:2008mg}. This technique has been
combined with the edge/endpoint method to achieve further
improvement~\cite{hybrid}.

In a seemingly parallel approach, several authors have studied mass
determination methods using the variable $m_{T2}$ \cite{mt2}, which is
sometimes called the {\it stransverse mass}. $m_{T2}$ is defined event by
event as a function of the invisible particle mass. Its endpoint or
maximal value over many events, denoted by $m_{T2}^{\max}$, gives an
estimate of the mother particle's mass in the beginning of the decay
chain. When the invisible particle's mass is unknown, one has to use a
trial mass to calculate $m_{T2}$ and only obtains an estimate of the
mass difference. However, it has been shown in Ref.~\cite{kink} that
if the two mother particles decay through three-body decays to the
invisible particles, a ``kink'' occurs on the $m_{T2}^{\max}$ curve as
a function of the trial mass. The position of the kink is actually at
the true value of the invisible particle mass, which allows us to
simultaneously determine the masses of both the invisible particle and
its mother particle. A generalized study of the kink method is
available in Ref.~\cite{Barr:2007hy}.

The purpose of this paper is to clarify the relation between the two
 mass determination techniques, {\it i.e.}, the one using
kinematic constraints and the one using the variable $m_{T2}$. An
apparent difference between the two approaches is that the former uses
the 4-momenta of the visible particles, while the latter is defined
solely on the plane transverse to the beam direction. Nevertheless, due
to the lack of total momentum measurement in the beam direction, the
longitudinal momenta of the two invisible particles can be
arbitrarily chosen, offsetting some of the information obtained from the
visible particles' longitudinal momenta. As a consequence, $m_{T2}$ is
equivalent to the ``minimal'' kinematic constraints discussed below.

\begin{figure}
\begin{center}
 \includegraphics[width=0.4\textwidth]{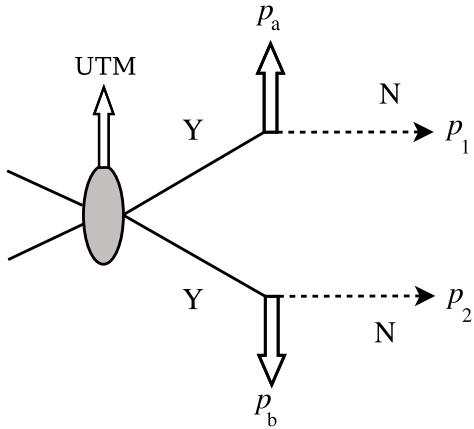}
\caption{\label{fig:mt2_define} An event with two invisible particles
  $N$, each from a decay of a heavy particle $Y$.} 
\end{center}
\end{figure}
We illustrate our definition of ``minimal'' constraints in
Fig.~\ref{fig:mt2_define}. Two mother particles
of the same mass, $m_Y$, each decays to a dark matter particle of mass
$m_N$, plus some visible particles, either directly
or through other on-shell particles. Since the two masses are
unknown, we have to assume some trial masses, denoted by $\mu_Y$ and
$\mu_N$. Upstream transverse momentum (UTM) can be present but it must
be known. The ``minimal'' constraints are then defined as the mass-shell
conditions from $\mu_Y$ and $\mu_N$, plus the constraint from the
measured $\slashchar p_T$. Obviously, for a given trial mass of the 
invisible particle $\mu_N$, the mother particle cannot be too light
otherwise we cannot obtain physical momenta for the invisible 
particles. As we will see, we can satisfy the minimal constraints and
obtain physical momenta if and only if $\mu_Y \ge m_{T2}$. We note
that this fact has been implicitly used in Ref.~\cite{m2c}, and we 
give a detailed account in this article. An important by-product of
our discussion is that we develop a new algorithm to calculate 
$m_{T2}$ which is 5--9 times as fast as the currently available program.

Since $m_{T2}$ corresponds to the minimal kinematic constraints,
it can serve as a starting point for mass determination with more
complicated topologies. The aforementioned ``kink'' method is an
example. The $m_{2C}$ variable defined in Ref.~\cite{m2c} is another
example, in which the authors combine $m_{T2}$ with the measurement of
the invariant mass distribution endpoint. We will give more examples
in this article.

The paper is organized as follows. In the next section, after
reviewing the definition of $m_{T2}$, we prove that it is equivalent to 
the minimal kinematic constraints. Deriving $m_{T2}$ from kinematic
constraints also provides us a fast algorithm for calculating
$m_{T2}$, which is presented in Section \ref{sec:calculate}. In Section
\ref{sec:mass_determine}, we discuss some mass determination methods from
our understanding of the relation between $m_{T2}$ and kinematic
constraints. A more general discussion of kinematic constraints and
conclusions are contained in Section \ref{sec:conclusion}.

\section{$m_{T2}$ from kinematic constraints}
\label{sec:mt2_constraints}

\subsection{The definition of $m_{T2}$}
\label{sec:mt2_define}

The definition of $m_{T2}$ is motivated from the transverse
mass $m_T$, which is defined for events with one invisible particle at
a hadron collider. In this case, the measured missing transverse
momentum is equal to the transverse momentum of the invisible
particle. $m_T$ has been used, for example, in the measurement 
of the $W$ boson mass in the decay $W\rightarrow \ell\nu$. Using the
notation of the $W$ decay, $m_T$ is defined by 
\begin{equation}
m_T^2=m_{\ell}^2+m_\nu^2+2(E_T^{\ell}E_T^\nu-{\bold p}_T^{\ell}\cdot
{\bold p}_T^\nu),
\end{equation}
where $m_\ell$ and $m_\nu$ are respectively the masses of the lepton
and the neutrino,  and ${\bold p}_T^{\ell}$ and ${\bold p}_T^\nu$ are
their transverse momenta. The beams are chosen to be along the $z$
direction, therefore ${\bold p_T}=(p_x,p_y)$. $E_T^{\ell}$, $E_T^\nu$
are transverse energies defined by
\begin{equation}
E_T^{\ell}=\sqrt{m_{\ell}^2+|{\bold p}_T^{\ell}|^2},
\  \ E_T^{\nu}=\sqrt{m_{\nu}^2+|{\bold p}_T^{\nu}|^2}.
\end{equation}
For convenience, we use $\alpha$ to denote the 2+1 dimensional
momentum, $\alpha=(E_{T},{\bold p}_T)$, while the 4-momentum is
denoted by $p$. In the 2+1 dimensional notation, the transverse
mass is given by
\begin{equation}
m_{T}^2=(\alpha_\ell+\alpha_\nu)^2. \label{eq:mt_define}
\end{equation} 
The following relation is always satisfied for the physical momenta
$p^{\ell}$, $p^\nu$ and the corresponding 2+1 dimensional momenta:
\begin{equation}
(p_{\ell}+p_{\nu})^2\ge(\alpha_\ell+\alpha_\nu)^2. \label{eq:inv_ge}
\end{equation}
The equality holds if and only if $\ell$ and $\nu$ have the same
rapidity, which is given by
\begin{equation}
\eta=\frac12\ln\left(\frac{E+p_z}{E-p_z}\right).
\end{equation}

When the event contains two or more missing particles, we can no longer
calculate the transverse mass because the transverse momentum of the
individual missing particle is unknown. As mentioned in the
Introduction, a particular interesting case is that there are two
decay chains in the event, each ends with an invisible particle of
species $N$. We further assume that each decay chain also contains a
particle of species $Y$, decaying to the particle $N$ plus some
visible particles. This is illustrated in Fig.~\ref{fig:mt2_define},
where we have labeled the invisible particles as $1$ and $2$, and
summed the visible 4-momenta to $p_a$ and $p_b$ for the two decay
chains respectively. We will treat $a$ and $b$ as two particles whose
masses may vary from event to event. There can be other upstream
transverse momentum (UTM) from, for example, initial state radiation
or heavier particle decays. However, it is important that 1 and 2 are
the only invisible particles.

Comparing with the $W$ decay example, we see two difficulties
associated with the above decay chains. First, $m_N$, the mass of the
particle $N$ is {\it a priori} unknown. Second, only the sum of the
two invisible particles' transverse momenta is measured. These
difficulties motivated the authors of Ref.~\cite{mt2} to define a
quantity $m_{T2}$,  using a trial $N$ mass (denoted by $\mu_N$) and
minimizing over all possible partitions of the measured transverse
momentum:  
\begin{eqnarray}
m_{T2}^2(\mu_N)&\equiv& \min_{{\mathbf p}_T^1+{\mathbf p}_T^2=\mathbf{\slashchar
    p}_T}\left[\max \{ m_T^2({\mathbf p}_T^1,\,{\mathbf p}_T^a;\,\mu_N),\,
    m_T^2({\mathbf p}_T^2,\,{\mathbf
      p}_T^b;\,\mu_N)\}\right]\nonumber\\
&=&\min_{{\mathbf p}_T^1+{\mathbf p}_T^2=\mathbf{\slashchar
    p}_T}\left[\max \{(\alpha_1+\alpha_a)^2, (\alpha_2+\alpha_b)^2\}\right],\label{eq:mt2_define}
\end{eqnarray}  
where in the second line we have rewritten the transverse mass using
the 2+1 dimensional notation. By definition, $m_{T2}$ is an
event-by-event quantity depending on the trial mass
$\mu_N$. Therefore, strictly speaking it is not a variable, but a
function of $\mu_N$. For a given $\mu_N$, we can examine the $m_{T2}$
distribution for a large number of events, which in general has an end
point. As discussed in Ref.~\cite{mt2}, the
$m_{T2}$ end point gives the correct mass of the particle $Y$ when the
trial mass is equal to the true mass of the missing particle $N$,
$\mu_N = m_N$. We can therefore use $m_{T2}$ to determine $m_Y$ if
$m_N$ is known, analogous to the $W$ mass measurement. Moreover, it has
recently been shown \cite{kink} that, even if $m_N$ is unknown, in some
cases, when we plot the $m_{T2}$ endpoint as a function of the trial
mass $\mu_N$, there is a kink at $\mu_N = m_N$. Thus both $m_N$ and
$m_Y$ can be determined by studying the $m_{T2}$ distribution. 

We will discuss mass determination using $m_{T2}$ in Section
\ref{sec:mass_determine}. Before that, we first give an alternative 
definition of $m_{T2}$, using the concept of kinematic constraints.

\subsection{ $m_{T2}$ from minimal kinematic constraints}
\label{sec:constraints}
By kinematic constraints, we mean two kinds of constraints imposing on
the 4-momenta of the invisible particles: the mass shell constraints
and the measured missing transverse momentum constraints. Specifically,
for the event in Fig.~\ref{fig:mt2_define}, we can write down the
following equations: 
\begin{eqnarray}
p_1^2&=&p_2^2=\mu_N^2,\nonumber\\
(p_1+p_a)^2&=&(p_2+p_b)^2=\mu_Y^2,\nonumber\\
p_1^x+p_2^x&=&\slashchar p^x,\,p_1^y+p_2^y=\slashchar p^y, \label{eq:kinematics}
\end{eqnarray}
where $\mu_Y$ is a trial mass for the particle $Y$. We call this set
of constraints ``minimal'' because they correspond to the shortest decay
chains. Note that for a given set of $(\mu_N,\,\mu_Y)$, the system contains
only 6 equations, which are not enough for completely determining $p_1$ and
$p_2$. Nevertheless, Eqs.~(\ref{eq:kinematics}) still constrain the
possible $(\mu_N,\,\mu_Y)$. In particular, we will shortly see that
for a given $\mu_N$, 
Eqs.~(\ref{eq:kinematics}) can be satisfied for some physical momenta
$p_1$ and $p_2$ if and only if $\mu_Y>m_{T2}(\mu_N)$. Here, a momentum
is ``physical'' if all of its components are real and the energy
component is positive. In other words, $m_{T2}(\mu_N)$ can be defined
as the boundary of the consistent region on the $(\mu_N,\,\mu_Y)$
plane, subject to the minimal constraints in
Eqs.~(\ref{eq:kinematics}). This fact has been used in Ref.~\cite{m2c} 
but without a clear proof. 

First, it is easy to show that $\mu_Y$ cannot go below $m_{T2}$ for a
fixed $\mu_N$. For any $(\mu_N,\,\mu_Y)$ in the consistent mass
region, there exist physical $p_1$ and $p_2$ satisfying
Eqs.~(\ref{eq:kinematics}). On the other hand, from
Eq.~(\ref{eq:inv_ge}), we have
\begin{equation}
\mu_Y^2=(p_1+p_a)^2=(p_2+p_b)^2\ge\max\{(\alpha_1+\alpha_a)^2,\,(\alpha_2+\alpha_b)^2\}.
\end{equation}
By definition, $m_{T2}$ is the minimum of
$\max\{(\alpha_1+\alpha_a)^2,\,(\alpha_2+\alpha_b)^2\}$ over all
partitions of the missing transverse momentum. Therefore, we conclude
that $\mu_Y\ge m_{T2}(\mu_N)$.  

For the reverse direction we need to prove that for a given $\mu_N$, the point
$(\mu_N,\,m_{T2}(\mu_N))$ is indeed in the consistent mass region. By
the definition of $m_{T2}$, there exist physical 2+1 dimensional
momenta satisfying
\begin{eqnarray}
&&\alpha_1^2=\alpha_2^2=\mu_N^2,\nonumber\\
&&m_{T2}^2=(\alpha_1+\alpha_a)^2\ge(\alpha_2+\alpha_b)^2,\nonumber\\
&&p_{1}^x+p_{2}^x={\slashchar p}^x,\ \ p_{1}^y+p_{2}^y={\slashchar p}^y.\label{eq:3dconfig}
\end{eqnarray} 
Note that if $(\alpha_1+\alpha_a)^2<(\alpha_2+\alpha_b)^2$, we can
simply exchange the labels. Given $\alpha_1$ and $\alpha_2$, we can
arbitrarily choose $p_{1z}$ and $p_{2z}$ (or equivalently, the
rapidities $\eta_1,\, \eta_2$) of particles 1 and 2, and
Eqs.~(\ref{eq:3dconfig}) are still satisfied. In particular, we can
choose a $p_{1z}$ such that $\eta_1=\eta_a$. In this case we have
$(p_1+p_a)^2=(\alpha_1+\alpha_a)^2=m_{T2}^2$. As for the other decay
chain, if $(\alpha_2+\alpha_b)^2=m_{T2}^2$, we choose $\eta_2=\eta_b$;
if $(\alpha_2+\alpha_b)^2<m_{T2}^2$, we have $(p_2+p_b)^2<m_{T2}^2$
when $\eta_2=\eta_b$, and $(p_2+p_b)^2\rightarrow\infty$ when
$\eta_2\rightarrow \pm \infty$, as a result, there exists an $\eta_2$
such that $(p_2+p_b)^2=m_{T2}^2$. In this way we obtain physical
momenta $p_1$ and $p_2$ which satisfy Eqs.~(\ref{eq:kinematics}) with
$\mu_Y=m_{T2}(\mu_N)$. This concludes our proof.  

\subsection{Calculating $m_{T2}$}
\label{sec:calculate}
In the previous subsection, we have shown that $m_{T2}$ is the boundary
of the mass region consistent with the minimal kinematic
constraints. This provides us not only a way to understand
$m_{T2}$, but also an effective method of {\it calculating\/} it. 

We start by discussing how to determine if a mass pair $(\mu_N,\,\mu_Y)$
is consistent with the constraints in Eqs.~(\ref{eq:kinematics}). Note
that $m_{T2}$ is invariant under any independent longitudinal boosts
for the particles $a$ and $b$. This allows us to set $p_{a}^z$ and
$p_{b}^z$ to zero for convenience. We also assume $m_a>0$ and $m_b>0$
for the moment.  

We first consider the decay chain involving particles 1 and $a$. From
the mass shell constraints $p_1^2=\mu_N^2$ and
$(p_1+p_a)^2=\mu_Y^2$, we can express $E_1$ in terms of $p_{1}^{x}$ and $p_{1}^{y}$:
\begin{equation}
E_1= \frac{p_{a}^{x}}{E_a}p_{1}^{x} + \frac{p_{a}^{y}}{E_a}p_{1}^{y} +
\frac{\mu_Y^2-\mu_N^2-m_a^2}{2E_a}. 
\end{equation}
In order to have $p_1$ physical, we must have
\begin{equation}
-p_{1}^{z2}=-(E_1^2-p_{1}^{x2}-p_{1}^{y2}-\mu_N^2)\le0. \label{eq:ellipse} 
\end{equation}
Eq.~(\ref{eq:ellipse}) imposes a constraint on possible $p_{1}^{x}$ and
$p_{1}^{y}$. It is straightforward to show that the allowed
$(p_{1}^x,\,p_{1}^y)$ is the region enclosed by an ellipse. We  will
distinguish an ellipse and the region that it encloses by calling the
latter an ``elliptical region.'' The size of the ellipse depends on
$\mu_Y$ monotonically. In particular, it shrinks to zero when
$\mu_Y=\mu_N+m_a$, in which case all three particles have the same
velocity.  

\begin{figure}
\begin{center}
 \includegraphics[width=\textwidth]{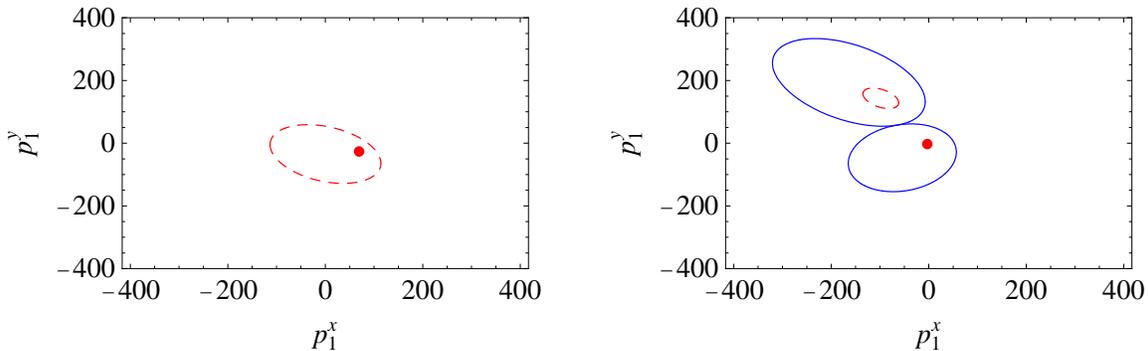}
\caption{\label{fig:ellipses} The unbalanced solution (left) and the
  balanced solution (right). The red (dashed) ellipse and the red
  point are the two ellipses when $\mu_Y=\mu_N+m_a$. For the
  unbalanced solution, the point is inside the red ellipse, and
  $m_{T2}=\mu_N+m_a$. For the balanced solution, the point is outside
  the red ellipse, and $m_{T2}$ is given when the two ellipses (solid
  blue) are tangent to 
each other. } 
\end{center}
\end{figure}

The other decay chain is completely analogous, and we obtain another
elliptical region for  $(p_{2}^x,\,p_{2}^y)$. However, the two decay
chains are related by the measured $\misspt$. Therefore, we can
eliminate $p_{2}^{x}$ and $p_{2}^{y}$ to put the second elliptical
region also on the $(p_{1}^x,\,p_{1}^y)$ plane. In order to satisfy
all the constraints, the two elliptical regions must overlap. Since
that the two ellipses both expand as we increase $\mu_Y$, $m_{T2}$
will be  given by the minimal $\mu_Y$ when the two elliptical regions
start to overlap. To proceed we need to distinguish two cases, which are
illustrated in Fig.~\ref{fig:ellipses} and discussed below. 

We assume $m_a \ge m_b$ for the invariant masses of $a$ and $b$
without loss of generality. We see that we must have $\mu_Y\ge
\mu_N+m_a$, otherwise the first ellipse vanishes. When
$\mu_Y=\mu_N+m_a$, the first ellipse becomes a point, while the second
ellipse has a finite size (or is also a point if $m_a=m_b$). If the
point (first ellipse of zero size) is within the second elliptical
region, then $m_{T2}$ is simply given by $m_{T2}=m_a+\mu_N$. This is
called the ``unbalanced configuration'' in Ref.~\cite{mtgen}.

The other possibility is that the point representing the zero-sized
first ellipse when $\mu_Y = \mu_N+m_a$ is outside the second
ellipse. In this case, we have to increase $\mu_Y$ until the two
elliptical regions overlap to obtain solutions. $m_{T2}$ is then given
by the value of $\mu_Y$ when the two ellipses are tangent to each
other. This is dubbed the ``balanced configuration'' \cite{mtgen}. 

Now it is clear how to calculate $m_{T2}$. For a given $\mu_N$, we
first check if the two ellipses give us an unbalanced configuration
when $\mu_Y = \mu_N+m_a$. If so, we have $m_{T2} =
\mu_N+m_a$. Otherwise, we need to look for the $\mu_Y$ when they are
tangent. The two ellipses are described by two quadratic equations,
which can be reduced to a univariate quartic equation. When the
ellipses are tangent, the quartic equation has degenerate roots and
therefore its discriminant vanishes. The discriminant is in general a
12th order polynomial function of $\mu_N^2$ and $\mu_Y^2$. This provides an
analytical relation between $\mu_N$ and $m_{T2}(\mu_N)$ which has not
be obtained in the literature before for the case of non-vanishing UTM. 
Although it would not be the most efficient way
to do the calculation, in principle one can
numerically solve the polynomial equation and obtain $m_{T2}$. Of
course, there can be more than one real solutions for the equation. One
should keep the smallest positive $\mu_Y$ as $m_{T2}$ since this is
the first time the two ellipses are tangent.  

It is convenient to use the discriminant when the UTM vanishes. In
this case, the equations are simplified so that the 
12th order equation is reduced to a 4th order one, for which analytical
solutions are available. This confirms the
existence of analytical solutions in the zero UTM case discussed in
Ref.~\cite{mtgen}. When UTM is nonzero, 
solving a 12th order equation is numerically slow and unstable. We
have developed a faster and more robust algorithm for calculating
$m_{T2}$, which is described in detail in Appendix
\ref{app:bisect}. The basic idea is that: we know that the two ellipses do
not intersect when  $\mu_Y^{\min} = \mu_N+m_a$ and we can also find a
$\mu_Y^{\max}$ when they do intersect by an educated guess. Then $m_{T2}$
must be within the interval $(\mu_Y^{\min},\mu_Y^{\max})$. Whether the two
ellipses intersect can be tested easily by the Sturm sequence~\cite{ellipse}. We
repeatedly bisect the interval while keeping the $m_{T2}$ within it,
until we reach the desired precision.   

In the above discussion, we have assumed that $m_{a,b}>0$. When either
 $m_a$ or $m_b$ (or both) vanishes, the corresponding ellipse becomes a
parabola, but the treatment remains the same.   

\section{Mass determination using $m_{T2}$}
\label{sec:mass_determine}
The simplest application of $m_{T2}$ is to determine $m_Y$ from the
$m_{T2}$ endpoint when $m_{N}$ is known. However, it is often the case
that $m_N$ is also unknown, and we want to determine both masses
simultaneously. The merit of $m_{T2}$ is that it corresponds to the
{\it minimal} constraints. Therefore it is always well-defined and
calculable, which may prove useful at the early stage of the LHC
\cite{inclusive_mt2}. If more information is available, we can develop
more complicated methods based on $m_{T2}$.

\begin{figure}
\begin{center}
 \includegraphics[width=0.4\textwidth]{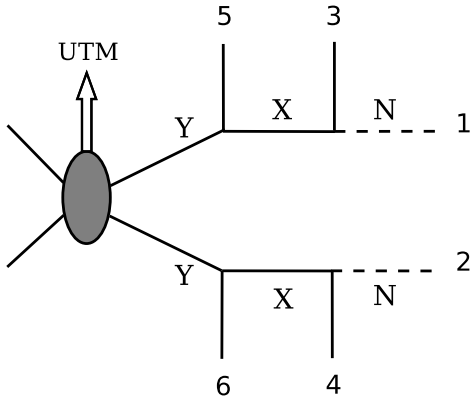}
\caption{\label{fig:2visible} The event topology with 2 visible
  particles per decay chain. The particle $X$ can be either on shell
  or off shell. } 
\end{center}
\end{figure}

If each decay chain in the events involves only a single two-body
decay, {\it i.e.}, $Y$ decays to $N$ plus a single
visible particle with fixed mass, it is impossible to determine both
masses from pure kinematics. If we consider two identical decay chains, 
the next simplest case is then that each decay chain contains
two visible particles, which is illustrated in
Fig.~\ref{fig:2visible}.  

When the particles in the two decay chains are identical, we can
divide the situations into four cases, depending on whether there is
significant UTM, 
and whether the intermediate particle $X$ is on shell or off
shell. The latter determines whether the particle $Y$ decays to $N$
through a three-body decay or two consecutive two-body decays. For the
two two-body decay cases, we would like to determine $m_X$ as well. The
cases with zero or negligible UTM are discussed in
Ref.~\cite{kink}. The authors pointed out that $m_{T2}^{\max}$,
the $m_{T2}$ endpoint value as a function of $\mu_N$, has a kink at
$\mu_N=m_N$. In practice, it is difficult to identify the kink due to
experimental smearing, but the formula for the $m_{T2}^{\max}$ curve
is known, which make it possible to fit the position of the kink.   

When the events have significant UTM, the situation is
different. UTM can come from initial state radiation or heavier
particle decays. For the latter, an example in MSSM is
the decay chain 
\begin{equation}
\tilde q\rightarrow q\tilde\chi_2^0\rightarrow q \ell\tilde\ell\rightarrow
q\ell\ell \tilde\chi_1^0,\label{eq:decaychain}
\end{equation}
where particles $\cntwo$, $\tilde\ell$ and $\cnone$ are identified
with $Y$, $X$ and $N$ respectively. The quark from squark decay can
be very energetic, providing large UTM to the system. In this case,
$m_{T2}^{\max}$ curve is different from the vanishing UTM case and an
analytical formula is in general unavailable. We focus on this case in
the following, taking the process in (\ref{eq:decaychain}) as an example.

We consider two mSUGRA points, one with
$m_{\tilde\ell}>m_{\tilde\chi_2^0}$ and the other one with
$m_{\tilde\ell}<m_{\tilde\chi_2^0}$, corresponding to the three-body
decay case and the two-body decay case respectively. The former is
chosen to be the same as 
the model P1 in Ref.~\cite{m2c} for comparison. The latter is the
Snowmass SUSY point SPS1a \cite{snowmass}.
\begin{enumerate}
\item $m_0=350\gev$, $m_{1/2}=180\gev$, $\tan\beta=20$,
  $\mbox{sign}(\mu)=+$, $A_0=0$;\\
$m_{\cntwo}=123.7\gev$, $m_{\tilde\ell_R}=358.6\gev$,
  $m_{\cnone}=70.4\gev$. 
\item $m_0=100\gev$, $m_{1/2}=250\gev$, $\tan\beta=10$,
  $\mbox{sign}(\mu)=+$, $A_0=-100$;\\
$m_{\cntwo}=181.0\gev$, $m_{\tilde\ell_R}=143.7\gev$,
  $m_{\cnone}=100.4\gev$. 
\end{enumerate} 
The spectra are calculated with SPheno \cite{spheno}. We have generated
1000 events for each case with MadGraph/MadEvent at the parton level.
In this paper, we will only consider the ideal case, {\it i.e.}, no
background, the particles are exactly on-shell, and there is no
experimental smearing or wrong combinatorics for the visible
particles. Nevertheless, we have avoided using features that is easily
lost after the above effects are included, such as a kink structure, and
expect our methods to be valid for realistic cases. A realistic study
is left for a future publication. 

\begin{figure}
\begin{center}
 \includegraphics[width=0.8\textwidth]{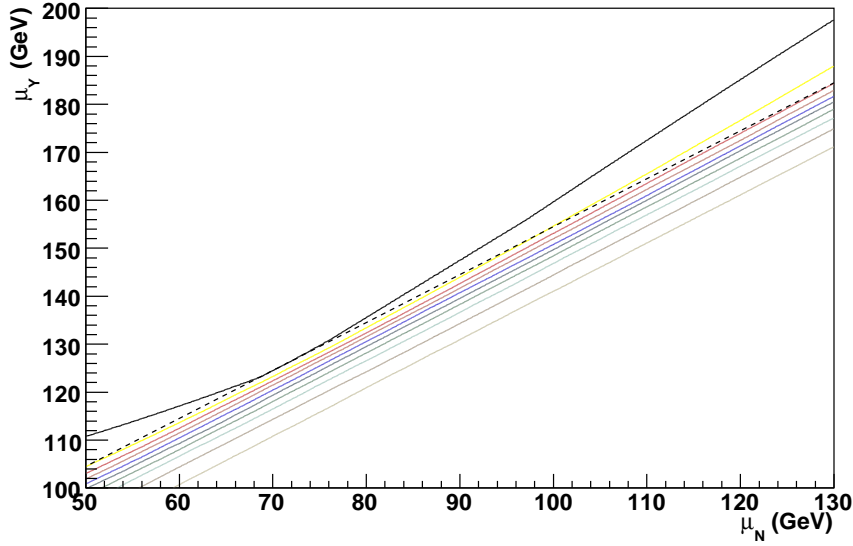}
\caption{\label{fig:contours_3body} Number of consistent events as a
  function of $(\mu_N,\,\mu_Y)$ for 1000 events, in the three-body
  decay case. The solid lines are
  the contours in 100 event intervals, beginning from 1000 for the top
  contour. The dashed line is the function $\mu_Y-\mu_N=M_{-}$.      } 
\end{center}
\end{figure}

We first discuss the three-body decay case. Again, it is illuminating
to think about $m_{T2}$ as kinematic
constraints. We have seen that for one event, the consistent mass region on
the $(\mu_N,\,\mu_Y)$ plane is above the $m_{T2}$
curve. Using this fact, for multiple events we can
easily count the number of events consistent with a given mass
point. Fig.~\ref{fig:contours_3body} is the contour plot for the
number of consistent events. In particular, above the
uppermost contour, which we identify as the $m_{T2}^{\max}$ curve, the
masses are consistent with all 1000 events. A kink structure in the
$m_{T2}^{\max}$ curve is visible. However it will be smeared by
experimental resolutions and we avoid using it. 

\begin{figure}
\begin{center}
 \includegraphics[width=0.5\textwidth]{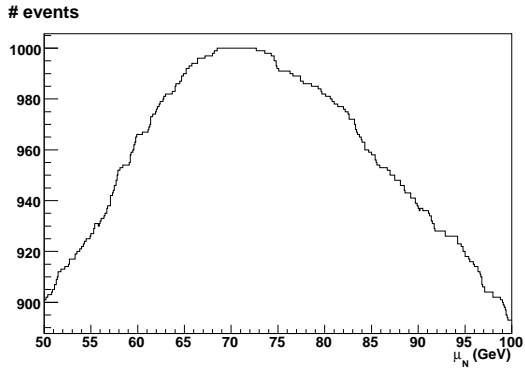}
\caption{\label{fig:nevents_3body} Number of consistent events along
  the line $\mu_Y-\mu_N=M_{-}$, in the three-body decay case. } 
\end{center}
\end{figure}

Similar to Ref.~\cite{m2c}, we can assume that the mass difference
$M_{-}=m_Y-m_N$ is already measured with good precision from the
endpoint of the dilepton invariant mass distribution. Indeed, due to
branching ratios, it is often the case that there are many more
dilepton events than four-lepton events and therefore the mass
difference can be measured much better. Drawing a line corresponding
to $\mu_Y-\mu_N=M_{-}$ on the $(\mu_N,\,\mu_Y)$ plane, we see that it
intersects some of the contours and touches the $m_{T2}^{\max}$ curve
only at $\mu_N=m_N$. We can draw the number of consistent events as a
function of $\mu_N$, along the line $\mu_Y-\mu_N=M_{-}$, as shown in
Fig.~\ref{fig:nevents_3body}. As expected, the number is maximized at
$\mu_N=m_N$. This is in some way equivalent to the approach in
Ref.~\cite{m2c}, where event-by-event lower and upper bounds for $m_Y$
are obtained by intersect the $m_{T2}$ curve with the line
$\mu_Y-\mu_N=M_{-}$. The distribution in Fig.~\ref{fig:nevents_3body}
is an integral of the upper and lower bound distributions of
Ref.~\cite{m2c}. Presenting in this way allows easy generalizations to other cases.
To minimize the statistical error, instead of simply
reading the maximum in Fig.~\ref{fig:nevents_3body}, we can also
fit the distribution to template distributions around the true $m_N$,
analogous to Ref.~\cite{m2c}. 

We now turn to the two-body decay case. In this case, the particle $X$
is on-shell, which gives us constraints in addition to
Eqs.~(\ref{eq:kinematics}) (where $p_a=p_3+p_5$ and $p_b=p_4+p_6$):
\begin{equation}
(p_1+p_3)^2=(p_2+p_4)^2=\mu_X^2,\label{eq:xconstraint}
\end{equation}
where $\mu_X$ is a trial mass for $X$. Unlike the minimal constraints,
the system given by combining Eqs.~(\ref{eq:kinematics}) and
(\ref{eq:xconstraint}) can be solved event by event to yield discrete
solutions for $p_1$ and $p_2$ (up to a four-fold
ambiguity)\footnote{This fact was first used to study $t\bar t$
events in the dilepton channel, see, for example, Ref.~\cite{ttbar}.}. Depending
on whether the solutions are physical, we can determine if an event is
consistent with a given set of masses $(\mu_N,\,\mu_X,\,\mu_Y)$. This fact
is used in Ref.~\cite{Cheng:2007xv} to determine all three masses by
examining the distribution of the number of consistent events. There, the
masses are obtained through a series of one-dimensional recursive
fits. In the following, we present a simplified method
utilizing the $m_{T2}^{\max}$ curve. The idea is that we can follow
the $m_{T2}^{\max}$ curve, which gives us a relation between $m_Y$ and
$m_N$, and count the number of consistent events. Depending on whether
we want to use measurement from the dilepton invariant mass
distribution like in the off-shell case, the method is slightly different.

\begin{figure}
\begin{center}
 \includegraphics[width=\textwidth]{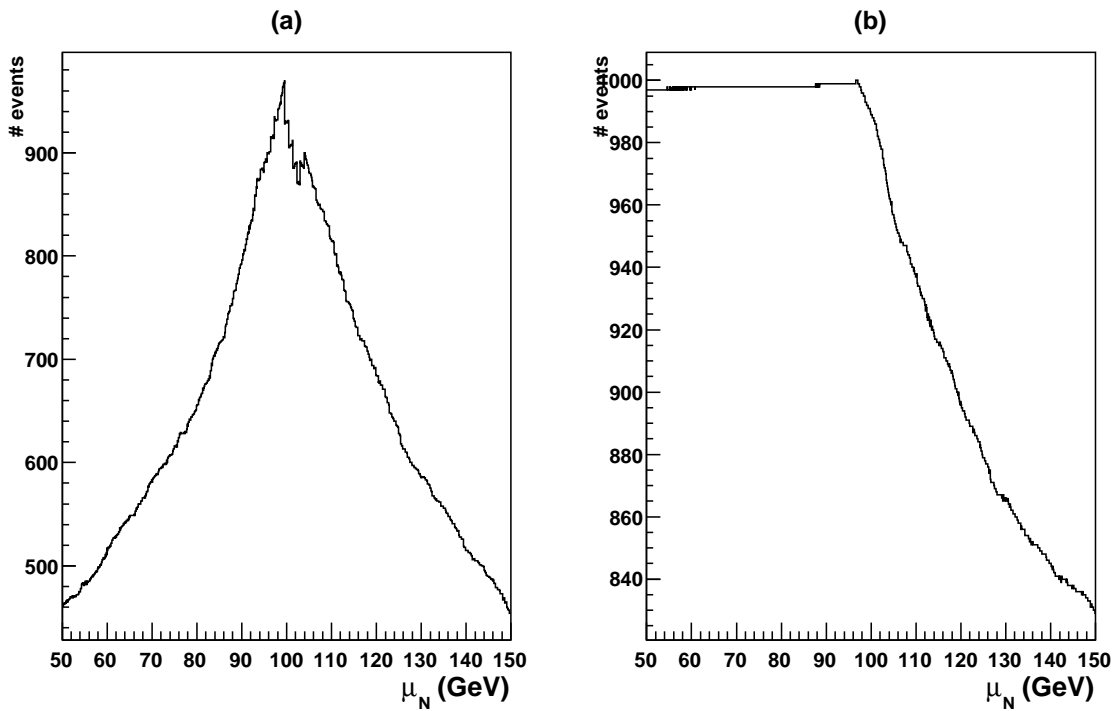}
\caption{\label{fig:nevents_2body} Number of consistent events
  distribution for $\mu_Y=m_{T2}^{\max}(\mu_N)$, in the two-body decay
  case. Left: $m_X$ is determined
  from the edge of the dilepton invariant mass distribution; right:
  $m_X$ is the value that maximize the number of consistent events. } 
\end{center}
\end{figure}
It is well known that for the decay chain in (\ref{eq:decaychain}) with
$\tilde\ell$ on shell, there
is a sharp edge in the dilepton invariant mass distribution at  
\begin{equation}
m_{\ell\ell}^2|_{edge}=\frac{(m_Y^2-m_X^2)(m_X^2-m_N^2)}{m_X^2}.\label{eq:edge}
\end{equation}
Assuming the edge position is measured with good precision, we obtain
a relation among the three trial masses. Together with the relation from
$m_{T2}^{\max}$ curve, $\mu_X$ and $\mu_Y$ are fixed for a given
$\mu_N$, up to a two-fold ambiguity from inversion of
Eq.~(\ref{eq:edge}). Then the number of events consistent with 
Eqs.~(\ref{eq:kinematics}) and (\ref{eq:xconstraint}), as a function
of $\mu_N$ is given in Fig.~\ref{fig:nevents_2body} (a). There is an
evident peak at $\mu_N=m_N$. 

The masses can be determined without using the measurement of the 
$m_{\ell\ell}$ edge. For each $\mu_N$, we first fix $\mu_Y$
by the $m_{T2}^{\max}$ function. Then we {\it vary} $\mu_X$ to
maximize the number of consistent events. This maximum number is shown
in Fig.~\ref{fig:nevents_2body} (b), as a function of
$\mu_N$\footnote{When $m_{T2}^{\max}(m_N)<m_Y$, which is always the case for
finite number of events, the number of solvable events at
$\mu_N=m_N$ drops much below the maximum number. Therefore, in
Fig.~\ref{fig:nevents_2body} (b) we have added to $\mu_Y$ a small
constant, $\mu_Y=m_{T2}^{\max}+2\gev$.}. Unlike the previous case, there is
not a peak structure, but the number of events drops sharply when
$\mu_N>m_N$, which can be used to estimate the masses.
 
\section{Discussion and conclusions}
\label{sec:conclusion}
We have demonstrated the relation between the $m_{T2}$ variable and
the kinematic constraints for events with two identical decay chains,
each of which ends up with one missing particle. The $m_{T2}^{\rm
  max}$ curve is equivalent to the boundary of the consistent region
in the mass space from the minimal kinematic constraints, where only
the mass shell conditions of the decaying mother particles and the
final missing particles, and the measured missing transverse momentum
constraint are imposed. In fact, it should not be surprising that many
different mass determination methods are closely related since they
are based on the same kinematics. Understanding their relations may
allow us to develop more effective and powerful ways for mass
determination either by finding new strategies or by combining various
approaches. Here we will try to give a general discussion of the mass
determination program based solely on kinematics.

For a given topology of new physics events, we can think of it as a
map between the ``mass'' space $\mathcal{M}$ and the ``observable''
space $\mathcal{O}$. The mass space is the space of the mass
parameters  of the new particles which appear on shell in those
events. The dimension is equal to the number of the unknown masses
that are to be determined. The observable space is the
multi-dimensional space of {\it all\/} independent kinematic
observables which are relevant for the mass determination in those
events. Basically, they are made of the momenta of the visible
particles (jets and leptons) from the decay chains, and the missing
transverse momentum. Therefore, each experimental event can be
represented by a point in the observable space. In principle we can
choose any basis for the observable space. It is convenient, however,
to choose combinations that are invariant under certain
transformations which do not alter the the connection between the mass
parameters and the kinematic observables. For example, if an event of
certain observable momenta can be produced by some mass parameters, a
boost along the beam axis can also be produced by the same mass
parameters because the momenta along the beam axis and the energies of
the initial partons in collision are unknown. They do not have to be
fully Lorentz-invariant as the measured missing transverse momentum
breaks the symmetry. For two decay chains in an event which is the
focus of most discussion, the allowed transformations are independent
boosts of the two decay chains along the beam axis and the rotation
around the beam axis, so it would be advantageous to choose the
observable combinations that are invariant under these
transformations. (If there is no UTM, one can also perform
back-to-back equal transverse boosts on the two chains as the two
mother particles have equal and opposite transverse momenta in this
case.)

\begin{figure}
\begin{center}
 \includegraphics[width=0.8\textwidth]{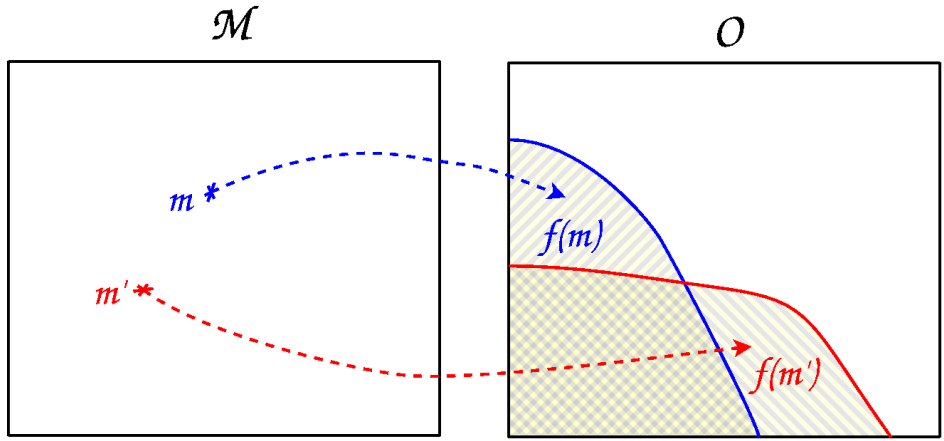}
\caption{\label{fig:map} The map between a point in the mass space and
the corresponding consistent region in the observable space.} 
\end{center}
\end{figure}
For each point $\mathbf{m}$ in the mass space, there is a
corresponding region $f(\mathbf{m}) \subset \mathcal{O}$ in the
observable space which is consistent with this mass point, {\it i.e.,}
$f(\mathbf{m})$ is made of all possible points in the observable space
that can be produced kinematically by the given mass parameters (see
Fig.~\ref{fig:map}). Assuming that there is a large enough set of
experimental events from this topology with the given mass parameters
at $\mathbf{m}$ and ignoring the issues such as experimental smearing
and backgrounds for the moment, the region $f(\mathbf{m})$ will be
populated by these experimental events. The relative weights and
densities of the experimental events within the region depend on
other details of the underlying theory such as the matrix elements. On
the other hand, the allowed region $f(\mathbf{m})$ solely depends on
the masses of the new particles. If $f(\mathbf{m})$ is unique for each
point $\mathbf{m}$, then in principle all the masses can be uniquely
determined given enough of experimental events. It is possible that
there are degeneracies such that different mass points map into the
same observable region, $f(\mathbf{m})=f(\mathbf{m}')$ for $\mathbf{m}
\neq \mathbf{m}'$, {\it e.g.,} the case of one step two-body decay on
each chain. In that case the masses cannot be uniquely determined
from kinematics alone and additional (model-dependent) information is
required. In general we expect $f(\mathbf{m})$ to be unique if the
dimension of the observable space is large enough. From the above
discussion, we see that the most important events for mass
determination are those which lie near the boundary of $f(\mathbf{m})$
as they determine the shape and the size of $f(\mathbf{m})$. The
edge/endpoint method can be viewed as a simple application of this idea
by projecting $f(\mathbf{m})$ down to a few one-dimensional subspaces
and extract the endpoints of $f(\mathbf{m})$ in these one-dimensional
subspaces. It is also evident that it does not fully utilize 
all the relevant information contained in the experimental events as
it only uses a few points on the boundary. In particular, in the case
of two visible particles in each decay chain it does not give enough
information to determine all masses, yet we know that the masses can
be determined by other methods. A generalization to look at the
boundary of the two-dimensional subspaces of $f(\mathbf{m})$ is
currently being studied~\cite{matchev}. It can potentially give a more
powerful method than the one-dimensional endpoint method. Ideally, one
would like to map out the whole boundary of $f(\mathbf{m})$ in the
high-dimensional observable space to get all the information contained
in the experimental events. However, dealing with the high-dimensional  
space could be technically quite difficult. 

\begin{figure}
\begin{center}
 \includegraphics[width=0.8\textwidth]{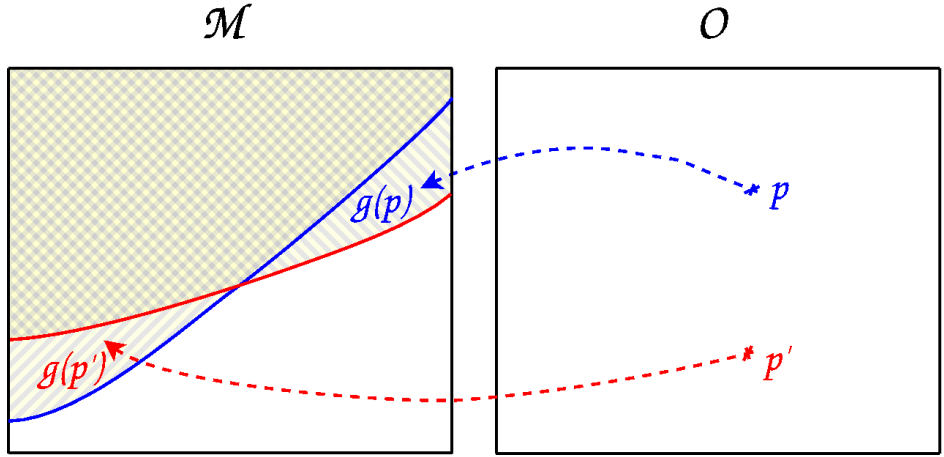}
\caption{\label{fig:inverse_map} The map between a point in the
observable space and the corresponding consistent region in the mass
space.}  
\end{center}
\end{figure}

The method of kinematic constraints can be considered as the inverse
map of the mass space and the observable space discussed above. Each
experimental signal event which is represented by a point $\mathbf{p}$
in the observable space $\mathcal{O}$ can define a region
$g(\mathbf{p})$ in the mass space $\mathcal{M}$ which is consistent
with that event (see Fig.~\ref{fig:inverse_map}). Obviously,
$\mathbf{m} \in g(\mathbf{p})$ if and only if $\mathbf{p} \in
f(\mathbf{m})$. The correct mass point must lie within the region
$g(\mathbf{p})$ assuming that the event is a valid signal event and no
experimental smearing. By combining many experimental events we can
find the intersection of all $g(\mathbf{p}_i)$ for all experimental
points $\mathbf{p}_i$, 
\begin{equation}
G(E) = \bigcap_{\mathbf{p}_i \in E} g(\mathbf{p}_i),
\end{equation}
where $E$ is the region in the observable space occupied by the
experimental events. If there are sufficiently many events, we expect
$E= f(\mathbf{m}_{\rm true})$ for the correct mass point
$\mathbf{m}_{\rm true}$. For certain event topologies, {\it e.g.,}
three on-shell two-body decays for each decay
chain~\cite{Cheng:2008mg}, the intersection region shrinks to a point 
after combining a finite number of events. Obviously it would
correspond to the correct mass point, $G(E)= \{\mathbf{m}_{\rm true}\}$, and
the mass determination is conceptually straightforward in this
case. Of course in practice, such determination will be complicated by
experimental smearing, backgrounds, and combinatorics. Non-trivial
techniques still need to be developed to resolve these issues in order
to demonstrate the viability and accuracies of mass determination in
this way. On the other hand, for many other topologies the region in
the mass space consistent with all experimental events $G(E)$ remains
finite. Na\"{i}vely one might think that this is a degeneracy and the
masses cannot be uniquely determined. However, a point in the mass
space $\mathbf{m}$ that is consistent with all events only implies
that $f(\mathbf{m}) \supset E$. For a generic but not the true mass
point in $G(E)$, $f(\mathbf{m})$ will be larger than $E$. Assuming
that the map between the two spaces is continuous, we then expect that
its immediate neighborhood points $\mathbf{m}'$ in many directions
will still be consistent with all events, $f(\mathbf{m}') \supset
E$. From this argument we see that any point lying in the middle of
$G(E)$ will not be the true mass point. For the true mass point, we
should have $f(\mathbf{m}_{\rm true})$ coincide exactly with $E$. Such
a point would have the least neighborhood points which are still
consistent with all events, or said in another way, it has the least
degrees of freedom to move while staying within $G(E)$. This tells us
that the true mass point should lie on the boundary of the consistent
mass region $G(E)$. In particular, if there is a sharp edge or a
``kink'' on the boundary of $G(E)$, it would be a good candidate for
the true mass point. This fact has been used to develop a new method
for mass determination in Ref.~\cite{Cheng:2007xv}. Now we see that
the method of the $m_{T2}$ kink~\cite{kink,Barr:2007hy} is another
example, as the $m_{T2}^{\rm max}$ curve is just the boundary of the
consistent region in the two-dimensional mass (sub)space based on the
minimal kinematic constraints. These methods effectively attempt to
match the whole region of $f(\mathbf{m})$ with the region of
experimental events $E$ as we hope to achieve in our discussion in the
previous paragraph. 

In the realistic situation, such a sharp edge or ``kink'' of the
consistent mass region can easily be washed out after the experimental
smearing and backgrounds are taken into account. Therefore it may not
be practical to directly search for the kink position. However,
understanding the structure of the consistent mass region from the
kinematic constraints allows us to develop strategies to recover the
kink location by combining various techniques that people have
developed. For example, it is well known that for collider signal
events with missing energies, the difficulty is to determine the
overall mass scale. The relative masses or mass differences usually
can be well constrained from the kinematic variables such as the
endpoints of invariant masses of visible particles. We can use those
kinematic variables to reduce the mass space down to a one (or low)
dimensional space which contains the true mass point. Then if we count
the number of consistent events as a function of the points along this
one dimensional space, it would, in the idealized case, exhibit a
sharp turning in number of consistent events at the true mass point
due to the ``kink'' nature of the consistent mass region near that
point. Even though the sharpness of the turning point will be reduced
by the experimental smearing and the presence of backgrounds, this
``turning'' feature is expected to survive as long as we have a
reasonable data set of the signal events, and we can fit for the
turning point to determine the overall mass scale. This was
illustrated in Section~\ref{sec:mass_determine}. 

In conclusion, we have clarified the relation between the $m_{T2}$
variable and the kinematic constraints for events with two decay
chains ending with invisible particles. $m_{T2}$ is a clever variable
which simply corresponds to the boundary of the allowed mass region
from the minimal kinematic constraints where only the constraints of
mass shell conditions of the mother particles and the missing
particles of the two decay chains, and the measured missing transverse
momentum are used. As a by-product, we also found a faster algorithm
to calculate $m_{T2}$ from the point of view of kinematic
constraints. These connections can also tell us how to develop new
ways by combining different existing methods to achieve the more
powerful and accurate mass determination for events with missing
energies. It will be extremely important for reconstructing the
underlying theory and verifying whether we have discovered the dark
matter particle if such new physics events with missing energies are
indeed found at the LHC.

\acknowledgments 
This work is supported in part by the U.S.~ Department of Energy grant
No.~DE-FG02-91ER40674. 

\appendix
\section{The bisection method for calculating $m_{T2}$}
\label{app:bisect}
We describe in this appendix the bisection algorithm for calculating
$m_{T2}$ for the balanced configuration. 

First, we need a method to quickly determine if two ellipses
intersect, {\it without} solving the quartic equation described in Section
\ref{sec:calculate}. This is done by calculating the Sturm
sequence for the quartic polynomial, which gives us the number of real
solutions for the quartic equation \cite{ellipse}. When the
number of real solutions are zero, either the two ellipses are outside
each other, or one completely contains the other one.

For the balanced configuration, the two ellipses are outside
each other for $\mu_Y^{\min}=\mu_N+\max\{m_a,
  m_b\}$. When we increase $\mu_Y$, both ellipses expand. It is easy
to see that they always intersect for $\mu_Y$ in some range. Thus, we
need to guess a point when they intersect. We do this by first finding
a $\mu_Y$ such that the two ellipses enclose a same point, for
example, the origin. In this case, either they intersect or one contains the
other one. If it is the former, we have found an intersecting point
which is taken as $\mu_Y^{\max}$. If it is the latter, which rarely
happens, we need to do a scan from $\mu_Y^{\min}$ to find the
intersecting point. 

After obtaining $\mu_Y^{\min}$ and $\mu_Y^{\max}$,  we bisect the
interval $(\mu_Y^{\min},\,\mu_Y^{\max})$ and check if the two
ellipses intersect at the middle point of the interval
$\mu_{Y}^{\mbox{\scriptsize mid}}=(\mu_Y^{\min}+\mu_Y^{\max})/2$. If
yes, we set the new  $\mu_Y^{\max}=\mu_{Y}^{\mbox{\scriptsize mid}}$;
otherwise, $\mu_Y^{\min}=\mu_{Y}^{\mbox{\scriptsize mid}}$. We repeat
this procedure until the size of the interval is smaller than the
precision we want.  

The algorithm has been implemented in c++ and available at
Ref.~\cite{code} or from the authors. The code 
has been tested for 250k events. These include 5 datasets with 50k
events each corresponding to the $t\bar t$ production in the dilepton
channel, and the two SUSY mass points discussed in Section
\ref{sec:mass_determine}. For the SUSY points, events with UTM (from
squark pair production and decay) and without UTM (from direct
$\tilde\chi_2^0$ pair production) are tested separately. The tests are
performed for a variety of trial masses $\mu_N$. The results have
been compared with Ref.~\cite{barr_lester_code}, showing good
agreement in the numerical values of $m_{T2}$: the possibility is
$O(10^{-5}\sim 10^{-4})$ for the two programs to yield values that differ by 1
GeV or more, and $O(10^{-4}\sim 10^{-3})$ for 0.1 GeV or more. For the
 events that give small differences, our code
is showing more accurate results, which can be verified in Mathematica
by examining when the two ellipses are tangent to each other. Our code
is also much faster (5--9 times as fast as
Ref.~\cite{barr_lester_code}), making it advantageous when $m_{T2}$
needs to be repeatedly calculated, for example, in evaluation of the
$m_{TGen}$ variable \cite{mtgen}.

\end{document}